\documentclass[12pt,a4paper]{article}
\usepackage{amsmath,amssymb,amsthm}
\usepackage[margin=1.0in]{geometry}
\usepackage{cite}
\usepackage{graphicx}
\usepackage{enumerate}
\usepackage{epsfig}
\usepackage{textcomp}
\usepackage{float}
\usepackage{bm}
\floatstyle{plaintop}
\restylefloat{table}
\allowdisplaybreaks
\usepackage[colorlinks=true
,urlcolor=blue
,anchorcolor=blue
,citecolor=blue
,filecolor=blue
,linkcolor=blue
,menucolor=blue
,linktocpage=true
,pdfproducer=medialab
,pdfa=true
]{hyperref}
\numberwithin{equation}{section}

\usepackage{adjustbox}
\usepackage{lipsum}
\usepackage{pbox}

\usepackage{graphicx}
\usepackage{subcaption}

\usepackage[font=footnotesize]{caption}

\def\d{\delta}

\def\m{\mu}
\def\n{\nu}

\def\L{\Lambda}

\def\be{\begin{equation}}
\def\ee{\end{equation}}
\def\bea{\begin{eqnarray}}
\def\eea{\end{eqnarray}}

\def\pa{\partial}

\def\lp{\left(}
\def\rp{\right)}

\def\nn{\nonumber}
\def\ie{{\it i.e., }}

\makeatletter
\renewcommand\section{\@startsection {section}{1}{\z@}%
	{-3.5ex \@plus -1ex \@minus -.2ex}
	{2.3ex \@plus.2ex}%
	{\normalfont\large\bfseries}}
\renewcommand\subsection{\@startsection{subsection}{2}{\z@}%
	{-3.25ex\@plus -1ex \@minus -.2ex}%
	{1.5ex \@plus .2ex}%
	{\normalfont\bfseries}}
\makeatother

\begin{document}

\begin{center}
\addtolength{\baselineskip}{.5mm}
\thispagestyle{empty}
\begin{flushright}
\end{flushright}

\vspace{20mm}

{\Large \bf Thermodynamics of charged rotating black strings \\ in extended phase space}
\\[15mm]
{Hamid R. Bakhtiarizadeh\footnote{h.bakhtiarizadeh@kgut.ac.ir}}
\\[5mm]
{\it Department of Nanotechnology, Graduate University of Advanced Technology,\\ Kerman, Iran}

\vspace{20mm}

{\bf  Abstract}
\end{center}

We investigate the thermodynamics of asymptotically Anti-de Sitter charged and rotating black strings in extended phase space, in which the cosmological constant is interpreted as thermodynamic pressure and the thermodynamic volume is defined as its conjugate. We find the thermodynamic volume, the internal energy, and the Smarr law. We study the thermal stability and show that some of the solutions have positive specific heat, which makes them thermodynamically stable. We find, for the first time, there is a critical point for charged solutions which occurs at the point of divergence of specific heat at constant pressure. This supports the existence of a second-order phase transition analogous to the liquid-gas critical point in Van der Waals fluids. We also study the maximal efficiency of a Penrose process and find that an extremal rotating black string can have an efficiency of up to 50\%. We also find the equation of state for uncharged solutions. By comparing with the liquid-gas system, we observe that there is not a critical behavior to coincide with those of the Van der Waals system. 

\vfill
\newpage


\section{Introduction}\label{int}

Recently, there has been significant interest in the field of black hole thermodynamics in extended phase space, where the cosmological constant and/or other coupling parameters are treated as thermodynamic variables \cite{Kastor:2009wy,Dolan:2010ha,Kubiznak:2012wp}. Although, the cosmological constant was traditionally considered a fixed parameter in the theory, however it could become a dynamical variable in gauged supergravity and string theories \cite{Cvetic:2010jb,Meessen:2022hcg}.

The black hole thermodynamics in extended phase space has offered a variety of physical implications and applications, for example, investigating black hole phase transitions and the emergence of a research direction known as black hole chemistry \cite{Kubiznak:2016qmn,Wei:2015iwa,Wei:2019uqg}. In the context of AdS/CFT, the holographic dual of the black hole thermodynamics in extended phase space on the CFT side has been widely explored. It also has been shown that the cosmological constant is not directly related to the CFT pressure but rather to the central charge \cite{Cong:2021fnf,Ahmed:2023snm}. The thermodynamics in extended phase space can also find applications in the study of holographic complexity, and other research fields \cite{AlBalushi:2020rqe,Gwak:2017kkt,Harlow:2022ich}. It is not only the cosmological constant that can be considered as a thermodynamic variable; but also other coupling parameters in higher curvature theories of gravity, just like those in Lovelock gravity, can also be regarded as thermodynamic variables \cite{Dutta:2022wbh,Kastor:2010gq,Frassino:2014pha,Dolan:2014vba,Sinamuli:2017rhp}. An elegant derivation of extended thermodynamics from the extended Iyer-Wald formalism has also been fulfilled in \cite{Xiao:2023lap}.

Specific interest developed in the thermodynamics of charged black holes in asymptotically AdS spacetimes \cite{Chamblin:1999tk,Chamblin:1999hg} after the leading work by Hawking and Page \cite{Hawking:1982dh}, who observed that there is a certain phase transition in the phase space of a Schwarzschild-AdS black hole. Since then, phase transitions and critical phenomena have been studied for more complicated backgrounds \cite{Cvetic:1999ne,Cvetic:1999rb}. The conventional phase space of a black hole consists only of entropy, temperature, charge and potential. Investigations of the critical behavior in the extended phase space seems more meaningful. With the extended phase space, the Smarr relation is satisfied in addition to the first law of thermodynamics, from which it is derived from scaling arguments. Furthermore, the resulting equation of state can be used for comparison with real world thermodynamic systems. This makes it evident that the phase space that should be considered is not the conventional one. Thermodynamic volume has been studied for a wide variety of black holes and is conjectured to satisfy the reverse isoperimetric inequality \cite{Cvetic:2010jb,Amo:2023bbo}. The critical behavior of charged and rotating AdS black holes in $ d $ spacetime dimensions, including effects from non-linear electrodynamics via the Born-Infeld action, in an extended phase space is also investigated in \cite{Gunasekaran:2012dq}. 

When the cosmological constant has been considered as a thermodynamic variable, some similarities with Van der Waals fluids will arise. For example, Joule-Thomson expansion which can occur in the region surrounding a black hole, where the strong gravitational field can create a pressure gradient that causes the gas to expand and cool. It is shown in \cite{Johnson:2014yja} that in theories of gravity where the cosmological constant is considered as a thermodynamic variable, it is natural to use black holes as heat engines. Similarities and differences between Van der Waals fluids and charged AdS black holes for the Joule–Thomson expansion are also investigated in \cite{Okcu:2016tgt}. The Joule–Thomson expansion for Kerr–AdS black holes in the extended phase space has been studied in \cite{Okcu:2017qgo}. In \cite{Nam:2019zyk} the heat engine and Joule–Thomson expansion for the charged AdS black hole in the context of nonlinear electrodynamics and massive gravity have been explored. The Joule-Thomson expansion in Einstein-Maxwell theory supplemented with the so-called quasitopological electromagnetism in the extended phase space thermodynamic approach has been studied in \cite{Barrientos:2022uit}. 

In this paper we are going to investigate the thermodynamic properties of the asymptotically AdS charged and rotating black strings in extended phase space. The asymptotically AdS charged rotating black strings were first introduced in \cite{Lemos:1994xp,Lemos:1995cm}. The thermodynamic properties of solutions in conventional phase space are also investigated in \cite{Dehghani:2002rr}. 

To avoid any confusion between the black strings (cylindrical black holes) found by Lemos with the black objects that appear in higher-dimensional general relativity like black rings \cite{Emparan:2001wn}, the following explanation sounds crucial. Lemos constructed black hole solutions in asymptotically AdS spacetimes where the event horizon topology is non-spherical. The metric solutions are static, and their existence is made possible due to the negative cosmological constant. In asymptotically flat spacetimes, such topologies are generally forbidden by topological censorship. So, Lemos's solutions show that when the cosmological constant is negative, event horizons of black holes can be ``extended" in one or more spatial directions, leading to black strings or black branes with translational symmetry; While, black rings are asymptotically flat solutions in five or more dimensions, where the event horizon is $ S^1\times  S^2 $ - a ring-like shape. Black rings are rotating, and rotation plays a critical role in balancing the ring against collapse. At high angular momentum, they resemble boosted black strings - long black objects with translational symmetry along a compact direction. These solutions emerge without needing a negative cosmological constant - unlike Lemos’s black strings. The existence of black rings violates uniqueness theorems from four-dimensional general relativity (no-hair theorem): in higher dimensions, multiple black objects can exist with the same mass and angular momentum. There is a conceptual link: both families show that black holes in higher dimensions or with non-trivial topology can differ radically from the four-dimensional Schwarzschild intuition. In both cases: 
\begin{enumerate}
	\item Event horizons can have non-spherical topologies.
	
	\item One can have solutions that resemble black strings or black branes (extended horizons).
	
	\item There are regimes where black rings resemble boosted black strings, drawing a parallel to Lemos's cylindrical solutions.
	
\end{enumerate}

However, the mechanism for obtaining these geometries differs - rotation and dimensionality in one case (black rings), versus cosmological constant and AdS boundary conditions in the other (black strings).

Moreover, it is possible to relate this four-dimensional black string solution to the BTZ three-dimensional solution \cite{Lemos:1995cp}. Generalizations of these solutions to higher dimensions which can be interpreted as black branes are introduced in \cite{Awad:2002cz} and their thermodynamics is examined in \cite{Dehghani:2002jh}. Solutions are also explored in the presence of Born-Infeld and power Maxwell invariant \cite{Hendi:2010kv} as well as logarithmic and exponential \cite{Hendi:2013mka} non-linear electrodynamics. A class of charged rotating black string solutions in four-dimensional Einstein-Maxwell-dilaton gravity with zero and Liouville-type potentials \cite{Dehghani:2004sa} as well as $ f(R) $-Maxwell theory \cite{Sheykhi:2013yga} is also constructed. Black string solutions in the context of mimetic gravity are explored in \cite{Sheykhi:2020fqf,Bakhtiarizadeh:2021pyo}. The critical behavior, phase transition and thermal stability of black strings in massive gravity is examined in \cite{Hendi:2020apr}.

The structure of our paper is arranged as follows. In the next section we review the metric and possible horizons of charged and rotating black strings. In Sec. \ref{therm}, we review the thermodynamic properties, the first law of thermodynamics in ordinary phase space. In Sec. \ref{vol}, we construct the theory of thermodynamics of charged and rotating black strings in extended phase space by considering the cosmological constant as thermodynamic pressure and introducing the extended first law of thermodynamics and Smarr law. We also find the thermodynamic volume, the mass and the internal energy as a function of thermodynamic variables. The thermal stability of solutions is also investigated in Sec. \ref{thermal}. In Sec. \ref{effi}, we explore the maximal efficiency of a Penrose process. In Sec. \ref{eqofst}, we extract the equation of state for uncharged black string and compare it with a liquid-gas system. Finally, Sec. \ref{conc} is devoted to conclusions.

\section{Black strings}\label{blackst}

The action for Einstein-Maxwell theory in four dimensions for asymptotically AdS spacetimes is\footnote{Throughout this paper we will work in units of $ G = 1 $, so one has to bear in mind that every quantity is expressed in Planck units.}
\be\label{action}
S=\frac{1}{16\pi}\int d^4 x \sqrt{-g}\lp R-2\L-F_{\mu \nu}F^{\mu \nu} \rp,
\ee
where $ R $ represents the Ricci scalar, $ \L = -3/\ell^2 $ is the negative cosmological constant of AdS space.\footnote{The asymptotically de-Sitter solutions can be obtained by taking $ \ell \to i\ell $.} Here, the field strength $ F_{\mu \nu} $ is given in terms of the vector potential $ A_{\m} $ by $ F_{\mu \nu}=2\pa_{[\mu} A_{\nu]} $. 

The metric of four-dimensional spacetime with cylindrical or toroidal horizons can be written as \cite{Lemos:1994xp,Lemos:1995cm} 
\bea\label{met}
ds^2=-f(r)\lp \Xi dt -a d\phi \rp ^2+\frac{1}{f(r)}dr^2+\frac{r^2}{\ell^4} \lp a dt -\Xi \ell^2 d\phi \rp ^2+\frac{r^2}{\ell^2}dz^2,
\eea
where
\be\label{sileq}
\Xi=\sqrt{1+\frac{a^2}{\ell^2}}.
\ee
Varying the action (\ref{action}) with respect to the metric $ g_{\m\n} $ and the vector field $ A_{\mu} $, we get the following field equations
\bea\label{eoms}
G_{\mu \nu}+\L g_{\mu \nu}= T_{\mu \nu},\nn\\
\nabla_{\mu} F^{\mu \nu} =0,
\eea
where
\be
T_{\mu \nu}=2 F_{\mu}{}^{ \rho}F_{\nu\rho}-\frac{1}{2}g_{\mu \nu}F_{\rho \lambda}F^{\rho \lambda}.
\ee
Notice that the metric is obtained after a boost along the $ (t,\phi) $ coordinates from the static black brane solution. The boost is only locally defined, so the global structure of the metric is different, but this explains why the metric takes this particular form and why it is not necessary to solve other components of the equations of motion, in particular the off-diagonal $ t\phi $ component of the field equations. The vector potential and nonzero components of electromagnetic field strength tensor are given by \cite{Lemos:1995cm}
\bea
&&A_{\m}=A_0(r)\lp \Xi \d_{\m}^{t}- a \d_{\m}^{\phi} \rp;\nn\\&&F_{tr}=-F_{rt}=-\Xi A'_0(r),F_{\phi r}=-F_{r\phi}=a A'_0(r).\label{nonvanF}
\eea
Here, a prime denotes a derivative with respect to $ r $. Having had the electromagnetic field strength tensor, it is not difficult to show that both $ t $ and $ \phi $ components of the Maxwell equation, lead to the same equation,
\be
r A''_0 +2 A'_0  =0.
\ee
This yields the following value for the gauge potential
\be
A_0(r)=-\frac{q}{r}.
\ee
Multiplying the $ zz $ component of the field equation by the factor $ r^2\Xi^2 $ and subtracting the resulting equation from
the $ \phi\phi $ component, we arrive at
\bea
r^2 \ell ^2 \lp r f'+f\rp-3 r^4+q^2 \ell ^2=0,
\eea
which leads to the following form for the metric function $ f(r) $,
\be\label{metfunc}
f(r)=\frac{r^2}{\ell^2}-\frac{m}{r}+\frac{q^2}{r^2}.
\ee
Here, the integration constants $ m $ and $ q $ are proportional to the mass and charge of black string and we call them the mass and charge parameters, respectively. The constants $ a $ and $ \ell $ have dimensions of length and can be interpreted as the rotation parameter and AdS radius, respectively. In the following, we are going to study the solutions with cylindrical symmetry. This implies that the spacetime admits a commutative two-dimensional Lie group $ G_2 $. The ranges of the time and radial coordinates are $ -\infty<t<\infty, 0\leq r<\infty $, and the topology of the horizon can be regarded as follows: \begin{enumerate}[(i)] \item the flat torus $ T^2 $ with topology $ S^1\times S^1 $ (\ie $ G_2=U(1) \times U(1) $) and the ranges $ 0\leq \phi<2\pi,0\leq z<2\pi \ell $, which describes a closed black string, \item the cylinder with topology $ \mathbb{R}\times S^1 $ (\ie $ G_2=\mathbb{R} \times U(1) $) and the ranges $ 0\leq \phi<2\pi,-\infty<z<\infty $, which describes a stationary black string, \item the infinite plane with topology $ \mathbb{R}^2 $ and the ranges $ -\infty<\phi<\infty,-\infty<z<\infty $, which does not rotate and can be interpreted as black hole.\end{enumerate}
We consider the topologies (i) and (ii) throughout the paper. 

\section{Thermodynamics}\label{therm}

The relevant thermodynamic potentials are given by
\bea\label{thermopot}
{\cal M}=\frac{1}{16 \pi \ell}\lp 3 \Xi^2-1\rp m,\qquad T=\frac{3 r_{+}^{4}-q^2 \ell^2}{4 \pi r_{+}^{3} \ell^2 \Xi},\qquad
{\cal S}=\frac{ r_{+}^{2}\Xi}{4\ell},\qquad
\Omega=\frac{a}{\Xi \ell^2},
\nn\\{\cal J}=\frac{3}{16 \pi \ell} \Xi a  m,\qquad\qquad\qquad \Phi=\frac{q}{r_+\Xi},\qquad\quad {\cal Q}=\frac{q \Xi}{4 \pi\ell},\qquad m=\frac{r_+^3}{\ell^2}+\frac{ q^2}{r_+},
\eea
where $ T $ is the Hawking temperature of the black string,  $ \Omega $ the angular velocity, $ \Phi $ the electrostatic potential difference between infinity and the horizon. Here also, $ {\cal M} $ is the mass, $ {\cal S} $ the entropy, $ {\cal J} $ the angular momentum and $ {\cal Q} $ the electric charge, per unit volume of black string horizon. The thermodynamic potentials per unit length of black string horizon are given in \cite{Dehghani:2002rr}. They also can be derived from \cite{Bakhtiarizadeh:2021vdo,Bakhtiarizadeh:2021hjr} by setting the coupling constant to zero. In the above, we rewrite them per unit volume of black string horizon.  

The event horizon is located at the largest root of the metric function (\ref{metfunc}) \ie $ f(r_+)=0 $. This equation leads to the mass parameter $ m $, given in Eq. (\ref{thermopot}), in terms of the horizon radius $ r_+ $, the charge paramete $ q $ and the AdS radius $ \ell $.
 
If we consider the mass per unit volume of black string horizon $ {\cal M} $ as a function of the entropy $ {\cal S} $, the charge $ {\cal Q} $, and the angular momentum $ {\cal J} $ per unit volume of black string horizon \ie $ {\cal M}\lp{\cal S},{\cal J},{\cal Q}\rp $, the first law of black hole thermodynamics,
\be\label{firstlaw}
d{\cal M} = Td{\cal S} + \Omega d{\cal J} + \Phi d{\cal Q},
\ee
is satisfied.

\section{Thermodynamic volume}\label{vol}

When a cosmological constant, $ \L $, is included, there is a natural candidate for thermodynamic pressure, 
\be\label{pressure}
P=-\frac{\L}{8\pi}=\frac{3}{8\pi\ell^2},
\ee
and it is proposed that the thermodynamic volume of the black string per unit horizon volume, $ {\cal V} $, be defined as the thermodynamic variable conjugate to $ P $. 

By allowing the pressure, the first law of black hole thermodynamics should then be modified to
\be\label{exfirstlaw}
d{\cal M} = Td{\cal S} + \Omega d{\cal J} + \Phi d{\cal Q}+{\cal V}dP.
\ee
The above extended first law is consistent with an integrated formula relating the black hole mass to other thermodynamic quantities known as Smarr relation:
\be\label{exsmarr}
0 = T {\cal S} + \Omega {\cal J} -2 P {\cal V}.
\ee
In writting the above result we have used the fact that under a change of length scale $ \lp{\cal M},{\cal S},{\cal J},{\cal Q},P\rp \propto \lp L^0,L^1,L^1,L^0,L^{-2}\rp$ and apply Euler’s theorem.\footnote{We also have used the fact that the volume of black string horizon in four dimensions, $ \omega_2 $, has dimension of length.}

The expression for mass $ {\cal M} $ in terms of thermodynamic variables $ ({\cal S}, P, {\cal J}, {\cal Q}) $ is
\bea\label{bsmass}
&&\!\!\!\!\!\!\!\!\!\!{\cal M} =\frac{1}{2^{3/4} \sqrt{3} \sqrt[4]{\pi } \sqrt[4]{P} \sqrt{\mathcal{S}} \left(32 \pi ^{3/2} \sqrt{2} \mathcal{J}^2 P^{3/2} \mathcal{S}+\varUpsilon \right)^{3/2}} \left[32 \pi ^{3/2} \sqrt{2} \mathcal{J}^2 P^{3/2} \mathcal{S} \varUpsilon +36864 P^4 \mathcal{S}^8\right.\nn\\&&\!\!\!\!\!\!\!\!\!\!\left.+55296 \pi  P^3 \mathcal{Q}^2 \mathcal{S}^6+32 \pi ^3 P \mathcal{S}^2 \left(64 \mathcal{J}^4 P^2+243 \mathcal{Q}^6\right)+31104 \pi ^2 P^2 \mathcal{Q}^4 \mathcal{S}^4+729 \pi ^4 \mathcal{Q}^8 \right],
\eea
where
\bea\label{varUpsilon}
&&\!\!\!\!\!\!\!\!\!\!\varUpsilon\equiv\left[110592 P^4 \mathcal{S}^8+165888 \pi  P^3 \mathcal{Q}^2 \mathcal{S}^6+32 \pi ^3 P \mathcal{S}^2 \left(64 \mathcal{J}^4 P^2+729 \mathcal{Q}^6\right)\right.\nn\\ &&\left.+93312 \pi ^2 P^2 \mathcal{Q}^4 \mathcal{S}^4+2187 \pi ^4 \mathcal{Q}^8\right]^{1/2}.
\eea
Differentiating with respect to $ P $ according to Eq. (\ref{exfirstlaw}) gives the thermodynamic volume per unit volume of black string horizon, in terms of thermodynamic variables as 
\bea\label{thermovolvari}
&&\!\!\!\!\!\!\!\!\!\!\mathcal {V}=\lp\frac{\pa {\cal {\cal M}}}{\pa P}\rp_{\cal S,\cal J,\cal Q} = \frac{\sqrt{32 \pi ^{3/2} \sqrt{2} \mathcal{J}^2 P^{3/2} \mathcal{S}+\varUpsilon }}{108\ 2^{3/4} \sqrt{3} \sqrt[4]{\pi } P^{5/4} \sqrt{\mathcal{S}} \left(8 P \mathcal{S}^2+3 \pi  \mathcal{Q}^2\right)^5}\times\nn\\&&\left[256 \pi ^{5/2} \sqrt{2} \mathcal{J}^2 P^{3/2} \mathcal{Q}^2 \mathcal{S} \varUpsilon +884736 P^5 \mathcal{S}^{10}+1216512 \pi  P^4 \mathcal{Q}^2 \mathcal{S}^8+580608 \pi ^2 P^3 \mathcal{Q}^4 \mathcal{S}^6\right.\nn\\&&\left.-8 \pi ^4 P \mathcal{Q}^2 \mathcal{S}^2 \left(2048 \mathcal{J}^4 P^2+729 \mathcal{Q}^6\right)+93312 \pi ^3 P^2 \mathcal{Q}^6 \mathcal{S}^4-2187 \pi ^5 \mathcal{Q}^{10}\right].
\eea
Transforming back to geometric variables $ (r_+, a ({\rm or}~\Xi), q, \ell) $, one finds
\be\label{thermovol}
{\cal V}=\frac{\Xi ^2 r_+^3}{4 \ell }+\frac{q^2 \ell\left(3 \Xi ^2-4\right) }{12 r_+}.
\ee
In terms of thermodynamic variables $ ({\cal S}, P, \cal J, \cal Q) $, the temperature is 
\bea\label{tempvari}
&&\!\!\!\!\!\!\!\!\!\! T =\lp\frac{\pa {\cal M}}{\pa {\cal S}}\rp_{P,\cal J,\cal Q}= -\frac{9 \sqrt{3} \left(\pi  \mathcal{Q}^2-8 P \mathcal{S}^2\right) \left(8 P \mathcal{S}^2+3 \pi  \mathcal{Q}^2\right)^3 \left(32 \pi ^{3/2} \sqrt{2} \mathcal{J}^2 P^{3/2} \mathcal{S} \varUpsilon +\varUpsilon ^2\right)}{2\ 2^{3/4} \sqrt[4]{\pi } \sqrt[4]{P} \mathcal{S}^{3/2} \varUpsilon  \left(32 \pi ^{3/2} \sqrt{2} \mathcal{J}^2 P^{3/2} \mathcal{S}+\varUpsilon \right)^{5/2}}, \nn\\
\eea
the angular velocity is
\bea\label{angvelvari}
\Omega=\lp\frac{\pa {\cal M}}{\pa {\cal J}}\rp_{{\cal S},P,\cal Q} = 
\frac{16\ 2^{3/4} \pi ^{5/4} \mathcal{J} P^{5/4} \sqrt{\mathcal{S}}}{\sqrt{3} \sqrt{32 \pi ^{3/2} \sqrt{2} \mathcal{J}^2 P^{3/2} \mathcal{S}+\varUpsilon }},
\eea
and the electric potential is
\bea\label{potvari}
&&\!\!\!\!\!\!\!\!\!\!\Phi =\lp\frac{\pa {\cal M}}{\pa \cal Q}\rp_{{\cal S}, P,\cal J}= \frac {\sqrt[4]{2} \pi ^{3/4} \mathcal{Q} \sqrt{32 \pi ^{3/2} \sqrt{2} \mathcal{J}^2 P^{3/2} \mathcal{S}+\varUpsilon }}{27 \sqrt{3} \sqrt[4]{P} \sqrt{\mathcal{S}} \left(8 P \mathcal{S}^2+3 \pi  \mathcal{Q}^2\right)^5}\times  \nn\\&&\!\!\!\!\!\!\!\!\!\!\left[-64 \sqrt{2} \pi ^{3/2} \mathcal{J}^2 P^{3/2} \mathcal{S} \varUpsilon +110592 P^4 \mathcal{S}^8+165888 \pi  P^3 \mathcal{Q}^2 \mathcal{S}^6\right.\nn\\&&\left.+32 \pi ^3 P \mathcal{S}^2 \left(128 \mathcal{J}^4 P^2+729 \mathcal{Q}^6\right)+93312 \pi ^2 P^2 \mathcal{Q}^4 \mathcal{S}^4+2187 \pi ^4 \mathcal{Q}^8\right].
\eea 
Transforming back to geometric variables $ (r_+, a ({\rm or}~\Xi), q, \ell) $, one finds the corresponding values in Eq. (\ref{thermopot}).

The internal energy per unit horizon volume of black string, $ {\cal U} $ is the Legendre transformation of the mass (enthalpy),
\be\label{intenergy}
{\cal U} = {\cal M} - P{\cal V},
\ee
where $ {\cal M} $ is a function of $ \cal S $, $ P $, $ \cal J $ and $ \cal Q $ while $ \cal U(\cal S, \cal V, \cal J, \cal Q) $ is a function of purely extensive variables. Then we get the usual form of the first law,
\be
d{\cal U} = Td{\cal S}+ \Omega d{\cal J} + \Phi d{\cal Q} -Pd{\cal V}.
\ee
Finding $ {\cal U} $ in terms of purely extensive variables $ (\cal S, \cal V, \cal J, \cal Q) $ is not possible because we can not solve Eq. (\ref{thermovolvari}) analytically to find $ P $  in terms of $ (\cal S, \cal V, \cal J, \cal Q) $. 

In the next section, it will be more convenient to use the variables $ {\cal S}, P, \cal J $ and $ \cal Q $, rather than $ \cal S, \cal V , \cal J $ and $ \cal Q $, in terms of which
\bea
&&{\cal U}=\frac {1}{4\ 2^{3/4} \sqrt{3} \sqrt[4]{\pi } \sqrt[4]{P} \sqrt{\mathcal{S}} \left(32 \pi ^{3/2} \sqrt{2} \mathcal{J}^2 P^{3/2} \mathcal{S}+\varUpsilon \right)^{3/2}}\times\nn\\&&\left[64 \pi ^{3/2} \sqrt{2} \mathcal{J}^2 P^{3/2} \mathcal{S} \varUpsilon +36864 P^4 \mathcal{S}^8+110592 \pi  P^3 \mathcal{Q}^2 \mathcal{S}^6\right.\nn\\&&\left.+128 \pi ^3 P \mathcal{S}^2 \left(32 \mathcal{J}^4 P^2+243 \mathcal{Q}^6\right)+93312 \pi ^2 P^2 \mathcal{Q}^4 \mathcal{S}^4+3645 \pi ^4 \mathcal{Q}^8\right].
\eea
In terms of geometrical variables,      
\bea
{\cal U}=\frac{r_+^4\left(3 \Xi ^2-2\right)+q^2 \ell ^2\left(3 \Xi^2+2\right)}{32 \pi r_+ \ell ^3}.
\eea

There is also a critical value of the black string charge parameter,
\be\label{extchargepara}
q_{\rm ext}=\frac{\sqrt{3} r_+^2}{\ell },
\ee
at which, the temperature vanishes, the horizon is degenerate and the black string is extremal. This leads to the following extremal value for mass parameter
\be\label{extmasspara}
m_{\rm ext}=\frac{4 r_+^3}{\ell ^2}.
\ee
Replacing the above two parameters into the thermodynamic potentials (\ref{thermopot}) and thermodynamic volume (\ref{thermovol}), one can easily find the extremal limit of thermodynamic variables. 

\section{Thermal stability}\label{thermal}

Here, we are going to investigate the thermal stability of the system in canonical ensemble. For an electrically neutral rotating black string, the specific heat at constant pressure per unit horizon volume is given by 
\bea
{\cal C}_{P}&=&T\lp \frac{\pa T}{\pa {\cal S}} \rp_{P,{\cal J}}^{-1}\nn\\&=&\frac{\mathcal{S} \left(\pi ^3 \mathcal{J}^4+54 P \mathcal{S}^6-9 \pi ^{3/2} \mathcal{J}^2 \sqrt{\pi ^3 \mathcal{J}^4+54 P \mathcal{S}^6}\right)}{27 P \mathcal{S}^6-40 \pi ^3 \mathcal{J}^4}.
\eea

\begin{figure}
	\centering
	\begin{subfigure}[b]{.496\linewidth}
		\includegraphics[width=\linewidth]{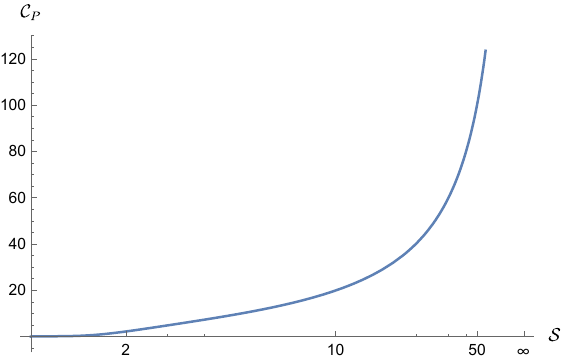}
		\setcounter{subfigure}{0}%
		\caption{}
	\end{subfigure}
	\begin{subfigure}[b]{.496\linewidth}
		\includegraphics[width=\linewidth]{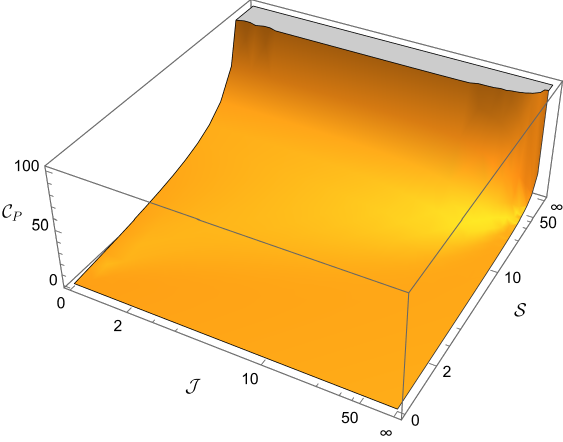}
		\caption{}
	\end{subfigure}
	\caption{The profiles of the specific heat per unit horizon volume at constant pressure for uncharged rotating black string as a function of (a) ${\cal S}$ with fixed values $ P=1 $ and ${\cal J}=1$, (b) ${\cal S}$ and ${\cal J}$ at fixed value $ P=1 $.}\label{fig1}
\end{figure}
\begin{figure}
	\centering
	\begin{subfigure}[b]{.496\linewidth}
		\includegraphics[width=\linewidth]{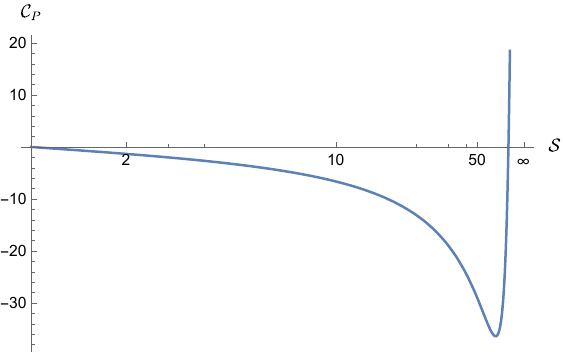}
		\setcounter{subfigure}{0}%
		\caption{}
	\end{subfigure}
	\begin{subfigure}[b]{.496\linewidth}
		\includegraphics[width=\linewidth]{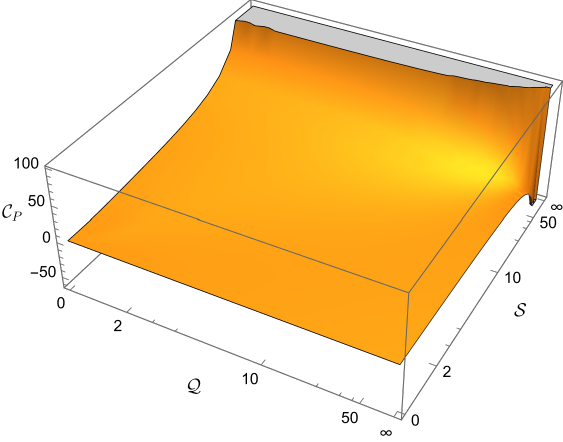}
		\caption{}
	\end{subfigure}
	\caption{The profiles of the specific heat per unit horizon volume at constant pressure for charged rotating black string as a function of (a) ${\cal S}$ with fixed values $ P=1 $, ${\cal J}=1$ and ${\cal Q}=250$ (b) ${\cal S}$ and ${\cal Q}$ at fixed value $ P=1 $ and ${\cal J}=1$.}\label{fig2}
\end{figure}
As illustrated in Fig. \ref{fig1}, the specific heat at constant pressure per unit horizon volume for uncharged rotating black string is positive and the solutions are thermodynamically stable. 
\newpage
In the case of charged solutions,
\bea
{\cal C}_{P}\!\!\!\!\!&=&\!\!\!\!\!T\lp \frac{\pa T}{\pa {\cal S}} \rp_{P,{\cal J},{\cal Q}}^{-1}\nn\\\!\!\!\!\!&=&\!\!\!\!\!-2 \mathcal{S} \left(\pi  \mathcal{Q}^2-8 P \mathcal{S}^2\right) \left(8 P \mathcal{S}^2+3 \pi  \mathcal{Q}^2\right) \left[-288 \sqrt{2} \pi ^{7/2} \mathcal{J}^2 P^{3/2} \mathcal{Q}^4 \mathcal{S} \varUpsilon \right.\nn\\&&\!\!\!\!\!\left.+4608 \pi ^{5/2} \sqrt{2} \mathcal{J}^2 P^{5/2} \mathcal{Q}^2 \mathcal{S}^3 \varUpsilon -18432 \sqrt{2} \pi ^{3/2} \mathcal{J}^2 P^{7/2} \mathcal{S}^5 \varUpsilon +7077888 P^6 \mathcal{S}^{12}\right.\nn\\&&\!\!\!\!\!\left.+15925248 \pi  P^5 \mathcal{Q}^2 \mathcal{S}^{10}+14929920 \pi ^2 P^4 \mathcal{Q}^4 \mathcal{S}^8+144 \pi ^5 P \mathcal{Q}^4 \mathcal{S}^2 \left(128 \mathcal{J}^4 P^2+2187 \mathcal{Q}^6\right)\right.\nn\\&&\!\!\!\!\!\left.+192 \pi ^4 P^2 \mathcal{Q}^2 \mathcal{S}^4 \left(512 \mathcal{J}^4 P^2+10935 \mathcal{Q}^6\right)+2048 \pi ^3 P^3 \mathcal{S}^6 \left(64 \mathcal{J}^4 P^2+3645 \mathcal{Q}^6\right)\right.\nn\\&&\!\!\!\!\!\left.+19683 \pi ^6 \mathcal{Q}^{12}\right]\left[452984832 P^8 \mathcal{S}^{16}+1358954496 \pi  P^7 \mathcal{Q}^2 \mathcal{S}^{14}+1783627776 \pi ^2 P^6 \mathcal{Q}^4 \mathcal{S}^{12}\right.\nn\\&&\!\!\!\!\!\left.+6912 \pi ^6 P^2 \mathcal{Q}^6 \mathcal{S}^4 \left(1024 \mathcal{J}^4 P^2+5103 \mathcal{Q}^6\right)+262144 \pi ^3 P^5 \mathcal{S}^{10} \left(5103 \mathcal{Q}^6-2560 \mathcal{J}^4 P^2\right)\right.\nn\\&&\!\!\!\!\!\left.+172032 \pi ^4 P^4 \mathcal{Q}^2 \mathcal{S}^8 \left(2048 \mathcal{J}^4 P^2+3645 \mathcal{Q}^6\right)+110592 \pi ^5 P^3 \mathcal{Q}^4 \mathcal{S}^6 \left(1701 \mathcal{Q}^6-512 \mathcal{J}^4 P^2\right)\right.\nn\\&&\!\!\!\!\!\left.+3779136 \pi ^7 P \mathcal{Q}^{14} \mathcal{S}^2+177147 \pi ^8 \mathcal{Q}^{16}\right]^{-1}.
\eea 
As can be seen from Fig. \ref{fig2}, the solutions are thermodynamically unstable unless at large values of $ {\cal S} $.

It also can be seen from Fig. \ref{fig2} that there is a critical point for charged solutions which occurs at the point of divergence of specific heat at constant pressure. This kind of divergence is a classical thermodynamic signal of a second-order phase transition for charged rotating black strings. These divergences typically occur at a critical point - analogous to the liquid-gas critical point in Van der Waals fluids. Note that, there is no first-order Hawking-Page phase transition here; because it is known that there is no Hawking-Page phase transition for a black object with a flat (zero curvature) horizon.

\section{The efficiency}\label{effi}

For non-zero $ \L $ we should expect the $ P d{\cal V} $ term to contribute to the mechanical energy that can be extracted from a black string, by a Penrose process for example. A negative $ \L $ and accordingly a positive pressure leads to a positive contribution to $ d{\cal U} $ if the black string shrinks, and the $ P d{\cal V} $ term reduces the amount of energy available for extraction as mechanical work $ {\cal W} $, with $ d{\cal W} = -d{\cal U} $, hence reducing the efficiency. Of course there are no pistons pushing against a gas for a black string, but a negative cosmological constant contributes a negative energy density to spacetime so a shrinking black string exposes negative energy, thus increasing the black string’s internal energy and decreasing the amount of energy available for mechanical work.

\begin{figure}
	\centering
	\begin{subfigure}[b]{.496\linewidth}
		\includegraphics[width=\linewidth]{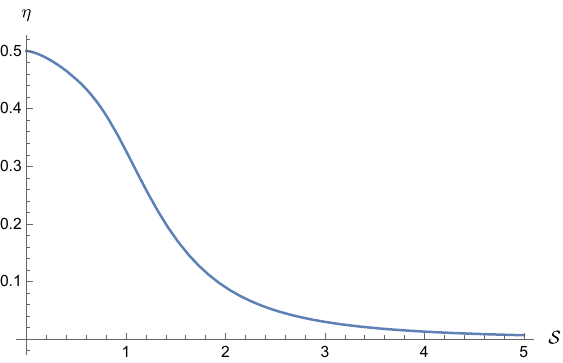}
		\setcounter{subfigure}{0}%
		\caption{}
	\end{subfigure}
	\begin{subfigure}[b]{.496\linewidth}
		\includegraphics[width=\linewidth]{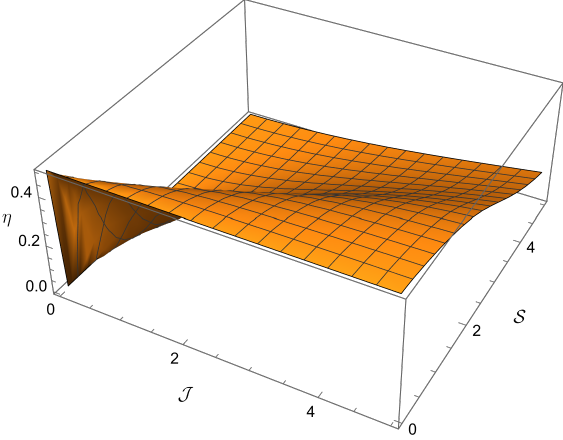}
		\caption{}
	\end{subfigure}
	\caption{The profiles of efficiency for uncharged solutions as a function of (a) ${\cal S}$ with fixed values $ P=1 $ and ${\cal J}=1$, (b) ${\cal S}$ and ${\cal J}$ at fixed value $ P=1 $.}\label{fig3}
\end{figure}

\begin{figure}
	\centering
	\begin{subfigure}[b]{.496\linewidth}
		\includegraphics[width=\linewidth]{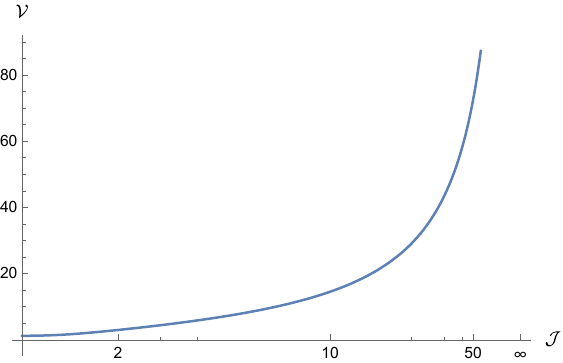}
		\setcounter{subfigure}{0}%
		\caption{}
	\end{subfigure}
	\begin{subfigure}[b]{.496\linewidth}
		\includegraphics[width=\linewidth]{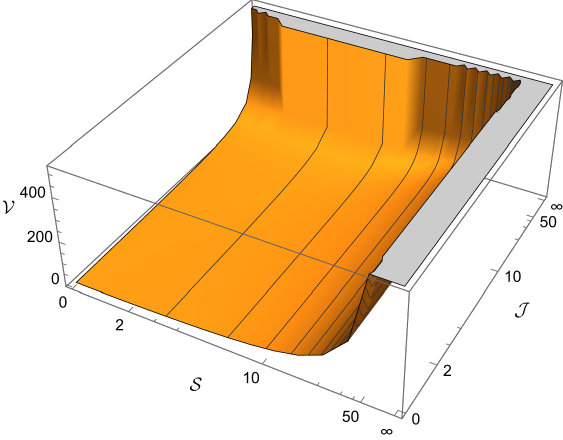}
		\caption{}
	\end{subfigure}
	\caption{The profiles of the thermodynamic volume for uncharged solutions as a function of (a) ${\cal J}$ with fixed values $ P=1 $ and ${\cal S}=1$, (b) ${\cal J}$ and ${\cal S}$ at fixed value $ P=1 $.}\label{fig4}
\end{figure}

\begin{figure}
	\centering
	\begin{subfigure}[b]{.496\linewidth}
		\includegraphics[width=\linewidth]{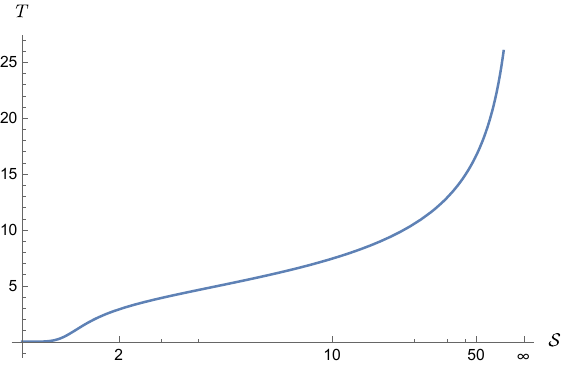}
		\setcounter{subfigure}{0}%
		\caption{}
	\end{subfigure}
	\begin{subfigure}[b]{.496\linewidth}
		\includegraphics[width=\linewidth]{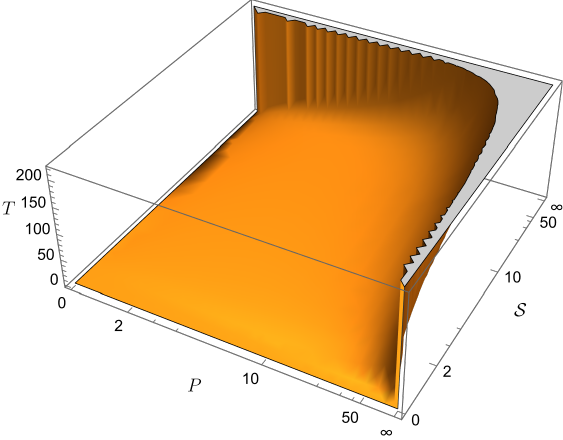}
		\caption{}
	\end{subfigure}
	\caption{The profiles of the Hawking temperature for uncharged solutions as a function of (a) ${\cal S}$ with fixed values $ P=1 $ and ${\cal J}=1$, (b) ${\cal S}$ and $P$ at fixed value ${\cal J}=1$.}\label{fig5}
\end{figure}

In an isentropic, isobaric process, an uncharged black string can yield mechanical work by decreasing its angular momentum. If $ {\cal J} $ is reduced from some finite value to zero, the efficiency is
\bea\label{uneffi}
\eta &=& \frac{{\cal U}({\cal J})-{\cal U}(0)}{{\cal M}}\nn\\&=&\frac{1}{12 \left(\pi ^3 \mathcal{J}^4+54 P \mathcal{S}^6+\pi ^{3/2} \mathcal{J}^2 \sqrt{\pi ^3 \mathcal{J}^4}+18 P \mathcal{S}^6\right)}\times\nn\\&&\left[6 \pi ^3 \mathcal{J}^4+54 P \mathcal{S}^6+6 \pi ^{3/2} \mathcal{J}^2 \sqrt{\pi ^3 \mathcal{J}^4}\right.\nn\\&& \left.-\sqrt[4]{2} 3^{3/4} \sqrt[4]{P} \mathcal{S}^{3/2} \left(\pi ^{3/2} \mathcal{J}^2+\sqrt{\pi ^3 \mathcal{J}^4+54 P \mathcal{S}^6}\right)^{3/2}\right].
\eea
In terms of geometrical variables,
\be\label{ungeoeffi}
\eta =\frac{\Xi ^{3/2}-3 \Xi ^2+2}{2-6 \Xi ^2}.
\ee 
There is a minimum value of $ \Xi $ for which $ \eta=0 $ and it occurs when $ \Xi=1 $ (or equivalently $ a=0 $), which means the black string does not possess any rotation. The greatest efficiency is for extremal black strings, when the Hawking tempereture vanishes. For a given $ \mathcal{S} $, $ P $ and $ \mathcal{J} $, there is a maximal value of $ \eta $ determined by demanding that $ T = 0 $, which requires $ {\cal S}=0 $. Vanishing the entropy in Eq. (\ref{uneffi}) leads to $ \eta = 1/2 $, as illustrated in Fig. \ref{fig3}.
It also can be seen that the thermodynamic volume is a monotonically increasing function of $ {\cal J} $, hence the volume decreases as $ {\cal J} $ is lowered, keeping the area constant, as illustrated in Fig. \ref{fig4}. As can be seen from Fig. \ref{fig5}, the black string also heats up during the process. 
\begin{figure}
	\centering
	\begin{subfigure}[b]{.496\linewidth}
		\includegraphics[width=\linewidth]{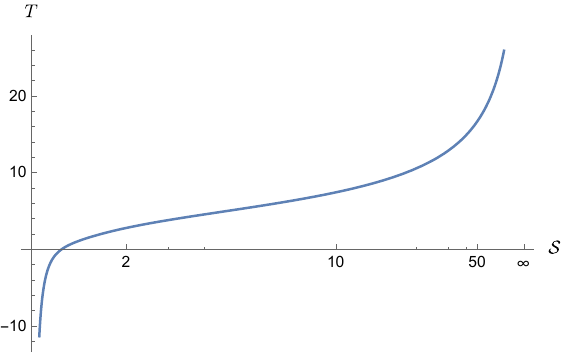}
		\setcounter{subfigure}{0}%
		\caption{}
	\end{subfigure}
	\begin{subfigure}[b]{.496\linewidth}
		\includegraphics[width=\linewidth]{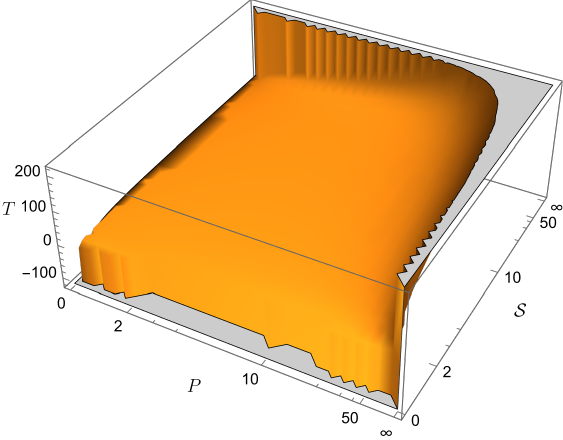}
		\caption{}
	\end{subfigure}
	\caption{The profiles of the Hawking temperature for charged solutions as a function of (a) ${\cal S}$ with fixed values $ P=1 $, ${\cal J}=1$ and ${\cal Q}=1$, (b) ${\cal S}$ and $P$ at fixed value ${\cal J}=1$ and ${\cal Q}=1$.}\label{fig6}
\end{figure}

\begin{figure}
	\centering
		\includegraphics{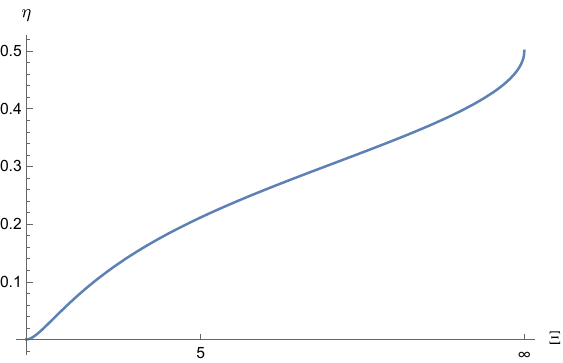}
	\caption{The profile of efficiency for charged solutions as a function of $\Xi$.}\label{fig7}
\end{figure}

When the black string is electrically charged and $ {\cal J} $ and $ {\cal Q} $ are both non-zero initially and are both decreased to zero, the efficiency is given by a complicated function of $ \cal S $, $ P $, $ \cal J $ and $ \cal Q $. But, in terms of geometrical variables, it takes the following simple form,  
\be\label{chgeoeffi}
\eta =\frac{\left(-5 \Xi ^{3/2}+3 \Xi ^2+2\right) q^2 \ell ^2+\left(-\Xi ^{3/2}+3 \Xi ^2-2\right) r_+^4}{2 \left(3 \Xi ^2-1\right) \left(q^2 \ell ^2+r_+^4\right)}.
\ee 
The greatest efficiency is for charged extremal black strings, for which one can replace the charge parameter $ q $ in the above equation with its extremal value given by Eq. (\ref{extchargepara}). Doing this gives
\be
\eta=\frac{1-2 \Xi ^{3/2}}{3 \Xi ^2-1}+\frac{1}{2}.
\ee
As illustrated in Fig. \ref{fig7}, it can be easily found that the maximum value of $ \eta $ is $ 1/2 $ which occurs for asymptotic value of $ \Xi $. It also can be seen from Fig. \ref{fig6} that in the case of charged solutions the black string heats up, as in uncharged case.

\section{Equation of state}\label{eqofst}

\begin{figure}
	\centering
	\begin{subfigure}[b]{.496\linewidth}
		\includegraphics[width=\linewidth]{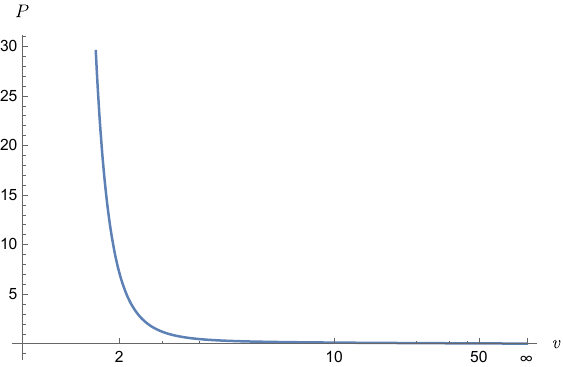}
		\setcounter{subfigure}{0}%
		\caption{}
	\end{subfigure}
	\begin{subfigure}[b]{.496\linewidth}
		\includegraphics[width=\linewidth]{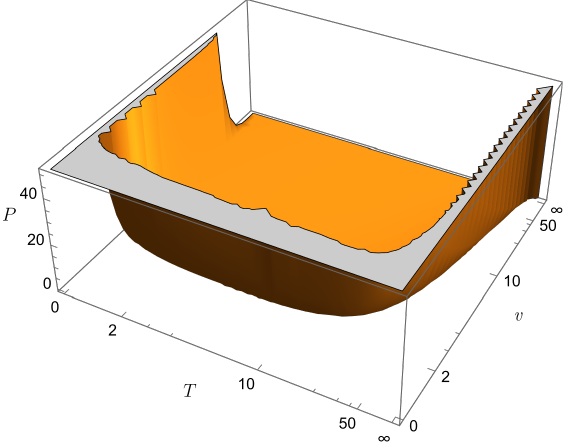}
		\caption{}
	\end{subfigure}
	\caption{The profiles of $ P $ as a function of (a) $v$ for uncharged rotating black string with fixed values $ T=1 $ and ${\cal J}=1$, (b) $v$ and $T$ at fixed value $ {\cal J}=1 $.}\label{fig8}
\end{figure}

When $ {\cal Q}=0 $, one can easily find the internal energy per unit horizon volume $ {\cal U} $ as a function of purely extensive variables $ ({\cal S},{\cal V},{\cal J}) $. To do so, we set $ {\cal Q}=0 $ in Eq. (\ref{thermovolvari}) for thermodynamic volume and solve it in terms of pressure, the result is 
\be\label{unchpressure}
P=\frac{2 \pi \mathcal{J}^2}{3  \mathcal{V}^2}+\frac{6 \mathcal{S}^6}{\pi \mathcal{V}^4}.
\ee
By setting $ {\cal Q}=0 $ in Eq. (\ref{bsmass})  for mass and plugging the above formula for pressure into it, one finds  
\be\label{unchmass}
 \mathcal{M}= \frac{4 \pi  \mathcal{J}^2}{3 \mathcal{V}}+\frac{8 \mathcal{S}^6}{\pi  \mathcal{V}^3}.
 \ee
Replacing Eqs. (\ref{unchpressure}) and (\ref{unchmass}) into Eq. (\ref{intenergy}), yields the following value for internal energy per unit horizon volume
\be\label{unchintenergy}
{\cal U}=\frac{2 \pi  \mathcal{J}^2}{3 \mathcal{V}}+\frac{2\mathcal{S}^6}{\pi \mathcal{V}^3}.
\ee
with temperature
\be\label{unchtemperature}
T=\lp\frac{\pa {\cal U}}{\pa \cal S}\rp_{{\cal V},\cal J}=\frac{12 \mathcal{S}^5}{\pi \mathcal{V}^3}.
\ee
The equation of state, in the form of the relation between the pressure, the temperature and the thermodynamic volume per unit horizon volume, can be obtained by eliminating $ \mathcal{S} $ between Eqs. (\ref{unchpressure}) and (\ref{unchtemperature}),  
\be\label{eqofstate}
P=\frac{2 \pi  \mathcal{J}^2}{3 \mathcal{V}^2}+\frac{\sqrt[5]{\frac{\pi }{3}} T^{6/5}}{2\ 2^{2/5} \mathcal{V}^{2/5}}.
\ee
To make contact with the Van der Waals fluid in four dimensions, we should write the equation of state in terms of physical volume. To do so, we first set $ {\cal Q}=0 $ in Eq. (\ref{thermovolvari}) to find the thermodynamic volume for uncharged solutions. Then, we rewrite the result in terms of horizon radius by substituting the entropy from Eq. (\ref{thermopot}). The result reads
\be\label{unchthermovolvari}
\mathcal{V}=\frac{\sqrt{\frac{\pi }{6}} \left(\sqrt{16 \mathcal{J}^2+P^2 r_+^6}+P r_+^3\right)}{2 \sqrt{P}}.
\ee
By plugging Eq. (\ref{unchthermovolvari}) into Eq. (\ref{eqofstate}), one finds the pressure as
\bea
P &=& \frac{1}{2 \sqrt{3} r_+^5}\left[r_+^8 T^2+r_+^{20/3} T^{4/3} \sqrt[3]{12 \mathcal{J} \left(18 \mathcal{J}+\sqrt{324 \mathcal{J}^2+3 r_+^4 T^2}\right)+r_+^4 T^2}\right.\nn\\&&\qquad\qquad\left.+\frac{r_+^{28/3} T^{8/3}}{\sqrt[3]{12 \mathcal{J} \left(18 \mathcal{J}+\sqrt{324 \mathcal{J}^2+3 r_+^4 T^2}\right)+r_+^4 T^2}}\right]^{1/2}.
\eea
Making an expansion in powers of the angular momentum up to order $ {\cal J}^2 $, one finds,
\be\label{eqofstateho}
P=\frac{T}{2 r_+}+\frac{4 \mathcal{J}^2}{r_+^5 T}+{\cal O}\lp {\cal J}^4 \rp.
\ee
Note that, in four dimensions
\be\label{plancklength}
l_{p}^2=\frac{\hbar G}{c^3}. 
\ee
To translate the ``geometric'' equation of state (\ref{eqofstateho}) to a physical one, note that the physical pressure and temperature are given by
\be\label{physpresstemp}
{\rm Press}=\frac{\hbar c}{l_{p}^2}P,\qquad {\rm Temp}=\frac{\hbar c}{k}T. 
\ee
When $ {\cal J}=0 $, multiplying (\ref{eqofstateho}) with $ \hbar c/l_{p}^2 $ we get
\bea
{\rm Press}&=&\frac{k{\rm Temp}}{2 r_+ l_{p}^2}.
\eea
Comparing with the Van der Waals equation, 
\be\label{VanderWaalseq}
\lp {\rm Press}+\frac{a}{v^2}\rp\lp v-b\rp=k {\rm Temp},
\ee
we conclude that, we should identify the specific volume $ v $ with
\be
v=2 r_+ l_{p}^2.
\ee
In geometric units, we have
\be
r_+=k v,\quad k=\frac{1}{2}.
\ee
By setting $ {\cal J}=0 $ in Eq. (\ref{unchthermovolvari}), one finds
\be\label{physicalvol}
v=2 \frac{\sqrt[6]{\frac{6}{\pi }} \sqrt[3]{\mathcal{V}}}{\sqrt[6]{P}}.
\ee
When $ {\cal J}\neq 0 $, by inserting (\ref{unchthermovolvari}) into (\ref{physicalvol}), we get   
\be
\mathit{v}=\frac{2^{2/3} \sqrt[3]{\sqrt{16 \mathcal{J}^2+P^2 r_+^6}+P r_+^3}}{\sqrt[3]{P}}.
\ee
Solving the above equation in terms of $ r_+ $ and substituting the result into Eq. (\ref{eqofstateho}), the equation of state can be rewritten as
\be
P=\frac{P^{2/3} T \mathit{v}}{\sqrt[3]{P^2 \mathit{v}^6-256 \mathcal{J}^2}}+\frac{128 \mathcal{J}^2 P^{10/3} \mathit{v}^5}{T \left(P^2 \mathit{v}^6-256 \mathcal{J}^2\right)^{5/3}}+{\cal O}\lp {\cal J}^4 \rp.
\ee
Making once again an expansion in powers of the angular momentum up to order $ {\cal J}^2 $, one finds,
\be
P=\frac{T}{\mathit{v}}+\frac{640 \mathcal{J}^2}{3 T \mathit{v}^5}+{\cal O}\lp {\cal J}^4 \rp.
\ee
The associated $ P-v $ diagram for an uncharged rotating black string is displayed in Fig. \ref{fig8}.
It can be seen there is no critical behavior and as a result no Hawking-Page phase transition. We also try to solve analytically the following two equations to obtain the critical point, 
\be
\frac{\pa P}{\pa v}=0,~\frac{\pa^2 P}{\pa v^2}=0,
\ee
and find that there is no root and accordingly there is no critical behaviour.

\section{Conclusion}\label{conc}

The thermodynamics of asymptotically AdS charged and rotating black strings in four-dimensional spacetime has been discussed in detail, with particular attention paid to the role of pressure and the volume. The cosmological constant is interpreted as a positive pressure and treated as a thermodynamic variable whose conjugate thermodynamic variable is a thermodynamic volume. 

The thermal stability of solutions has been analyzed in canonical ensemble and it has been observed that the uncharged solutions have positive specific heat at constant pressure and accordingly are thermodynamically stable while the charged ones are thermodynamically stable only for large values of entropy. 

By analyzing the diagram of specific heat at constant pressure versus entropy, it also has been found that there is a critical point for charged solutions which occurs at the point of divergence of specific heat at constant pressure. This means there is a second-order phase transition for charged rotating black strings just like what happens to the liquid-gas critical point in Van der Waals fluids.

In a Penrose process the volume decreases as the black string loses angular momentum and the $ Pd{\cal V} $ term in the first law reduces the amount of energy available to do work. Efficiencies of up to 50\% are theoretically achievable for a black string in asymptotically AdS spacetime.

The black string equation of state has been analysed in terms of pressure and physical volume. Non-zero angular momentum causes a rapid rise in pressure as the volume is reduced at constant temperature. We have found that there is no critical behavior and thus no Hawking-Page phase transition.

Let us close with some possible future explorations. It would be interesting to perform this kind of study in higher dimensions. The asymptotically AdS charged and rotating solutions in $ d $–dimensions have been presented in \cite{Awad:2002cz}, and their thermodynamics in ordinary phase space has been explored in \cite{Dehghani:2002jh}. These solutions, depending on their global identifications, have toroidal, planar or cylindrical horizons and can be interpreted as black holes, or black strings/branes. Finding the solutions in the presence of nonlinear electrodynamics \cite{Bakhtiarizadeh:2023mhk} is also an interesting feature.



\providecommand{\href}[2]{#2}\begingroup\raggedright
\endgroup
\end{document}